\documentclass[10pt]{article}
\usepackage{graphicx}
\usepackage{amsmath}

\def\Title#1{\begin{center} {\Large #1 } \end{center}}
\def\Author#1{\begin{center}{ \sc #1} \end{center}}
\def\Address#1{\begin{center}{ \it #1} \end{center}}

\newcommand\pubblock{\rightline{\begin{tabular}{l} Proceedings of the Fifth Annual LHCP\\ \pubnumber\\
         \pubdate  \end{tabular}}}

\newenvironment{Abstract}{\begin{quotation} \begin{center} 
             \large ABSTRACT \end{center}\bigskip 
      \begin{center}\begin{large}}{\end{large}\end{center} \end{quotation}}

\newenvironment{Presented}{\begin{quotation} \begin{center} 
             PRESENTED AT\end{center}\bigskip 
      \begin{center}\begin{large}}{\end{large}\end{center} \end{quotation}}

\def\beq{\begin{equation}}
\def\eeq#1{\label{#1}\end{equation}}
\def\eeqn{\end{equation}}

\def\beqa{\begin{eqnarray}}
\def\eeqa#1{\label{#1}\end{eqnarray}}
\def\eeqan{\end{eqnarray}}

\let\bar=\overbar

\def\Dslash{\not{\hbox{\kern-4pt $D$}}}
\def\dslash{\not{\hbox{\kern-2pt $\del$}}}

\def\msb{{\bar{\ssstyle M \kern -1pt S}}}

\usepackage{color}

\newcommand\pubnumber{ LHCb-PROC-2017-025 }

\newcommand\pubdate{\today}

\def\affiliation{
On behalf of the LHCb Collaboration, \\
Universit\`a di Bologna, Dipartimento di Fisica e Astronomia,\\
Istituto Nazionale di Fisica Nucleare - Sezione di Bologna,\\
viale Berti Pichat 6/2, Bologna (40127), Italy}

\usepackage{ifthen}
\newboolean{uprightparticles}
\setboolean{uprightparticles}{false} 
\usepackage{xspace} 
\usepackage{upgreek}

\def\Pnu         {\ensuremath{\nu}\xspace}
\def\neub       {{\ensuremath{\overline{\Pnu}}}\xspace}
\def\neutb      {{\ensuremath{\neub_\tau}}\xspace}

\def\neumb      {{\ensuremath{\neub_\mu}}\xspace}
\def\PB      {\ensuremath{B}\xspace}
\def\Bbar    {{\ensuremath{\kern 0.18em\overline{\kern -0.18em \PB}{}}}\xspace}
\def\PD      {\ensuremath{D}\xspace}                 
\def\Dbar    {{\kern 0.2em\overline{\kern -0.2em \PD}{}}\xspace}
\def\Dzb     {{\ensuremath{\Dbar{}^0}}\xspace}

\begin{document}

\large
\begin{titlepage}
\pubblock

\vfill
\Title{  TESTS OF LEPTON FLAVOUR UNIVERSALITY WITH SEMILEPTONIC DECAYS AT LHCb  }
\vfill

\Author{ FEDERICO BETTI  }
\Address{\affiliation}
\vfill
\begin{Abstract}

The observable $\mathcal{R} ( D^{(*)} ) = \mathcal{B}\left( B^{0}\to D^{(*)-} \tau^{+} \nu_{\tau} \right) / \mathcal{B}\left( B^{0}\to D^{(*)-} \mu^{+} \nu_{\mu} \right)$ is a probe for Lepton Universality violation, so it is sensitive to New Physics processes.
The current combination of the measurements of $\mathcal{R} ( D^{(*)} )$ differs from Standard Model predictions with a $4\sigma$ significance.
A measurement of $\mathcal{R} ( D^* )$ using three-prong $\tau$ decays has been performed at LHCb, resulting in $\mathcal{R}(D^*) = 0.285 \pm 0.019 (\text{stat}) \pm 0.025(\text{syst}) \pm 0.014 (\text{ext})$.
This value, combined with the LHCb result obtained with $\tau \to \mu \nu_\tau \neumb$ decays, gives ${\mathcal{R}}(D^*) = 0.306 \pm 0.016 (\text{stat}) \pm 0.022 (\text{syst})$, consistent with the world average and 2.1 standard deviations above the SM prediction.

\end{Abstract}
\vfill

\begin{Presented}
The Fifth Annual Conference\\
 on Large Hadron Collider Physics \\
Shanghai Jiao Tong University, Shanghai, China\\ 
May 15-20, 2017
\end{Presented}
\vfill
\end{titlepage}
\def\thefootnote{\fnsymbol{footnote}}
\setcounter{footnote}{0}
%

\normalsize 


\section{Introduction}

In the Standard Model (SM) of particle physics the electroweak couplings of the gauge bosons to the leptons are independent of their flavour, a property known as lepton universality (LU), so the observation of LU violation would be a clear signal of physics processes beyond the SM.

The branching fractions ratio:
\begin{equation}
	\mathcal{R} ( D^{(*)} ) = \frac{ \mathcal{B}\left( B^{0}\to D^{(*)-} \tau^{+} \nu_{\tau} \right) }{ \mathcal{B}\left( B^{0}\to D^{(*)-} \mu^{+} \nu_{\mu} \right) }
\label{eq:RDstar}
\end{equation}
represents a sensitive probe for LU violation.

The combination of the measurements of $\mathcal{R} ( D^{(*)} )$ already performed by BaBar~\cite{babar2013}, Belle~\cite{belle2015,belle2016_1,belle2016_2} and LHCb~\cite{lhcb2015} shows a discrepancy of about $4 \sigma$ with respect to the values of $\mathcal{R}( D^{(*)} )$ calculated within the SM~\cite{fajfer2012}.

\begin{figure}[htb]
	\centering
	\includegraphics[height=5in]{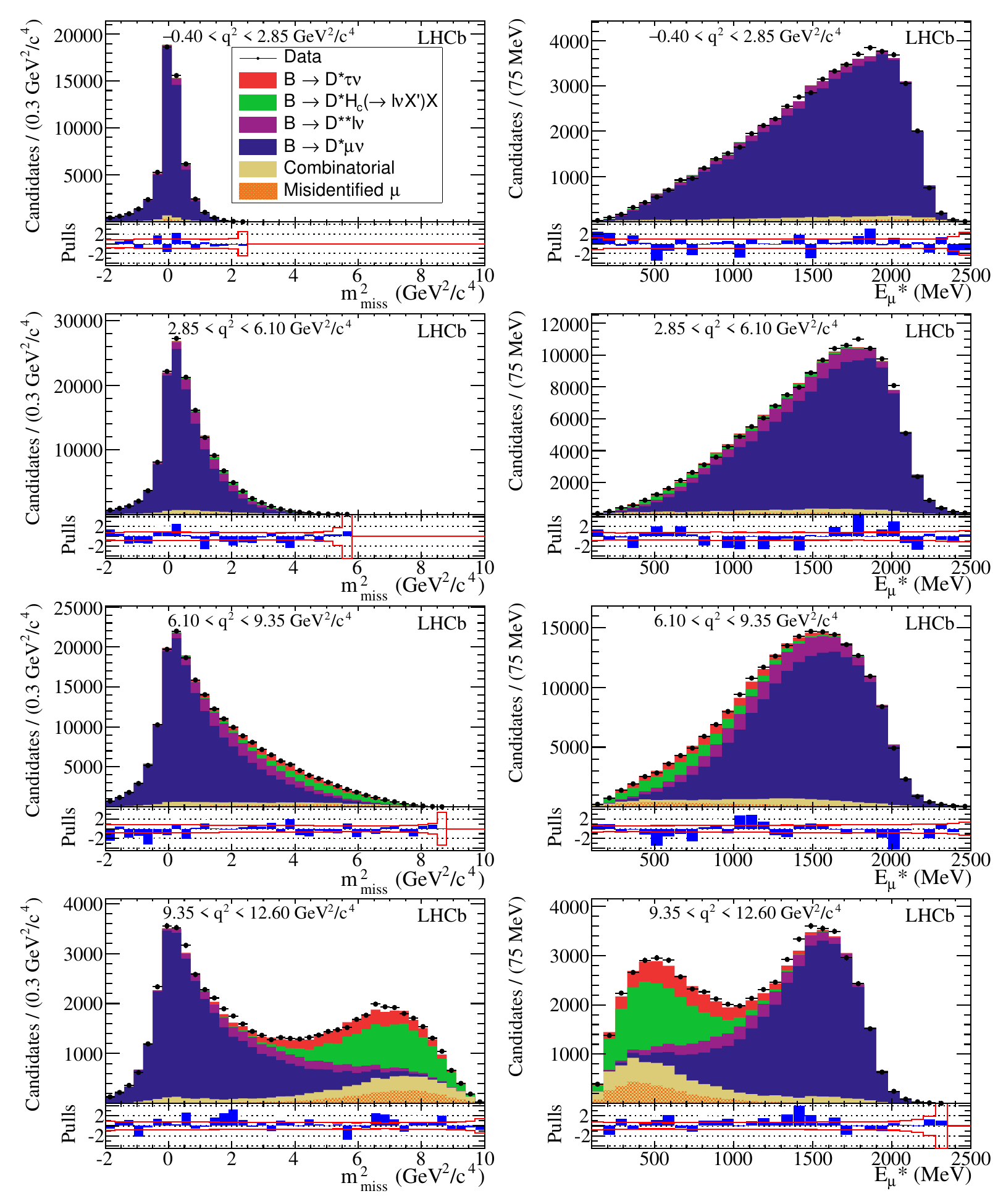}
	\caption{ Distributions of $m_{miss}^2$ and $E^*_\mu$ of the four $q^2$ bins of the signal data, overlaid with projections of the fit model with all normalization and shape parameters at their best-fit values.
	}
	\label{fig:fit_mu}
\end{figure}

The measurement performed by LHCb has been done reconstructing the $\tau$ lepton in the muonic decay mode $\tau \to \mu \nu_\tau \neumb$.
In this analysis three quantities, namely the transferred momentum $q^2$, the squared missing mass $m_{miss}^2$ and the muon energy in the $B$ rest frame $E^*_\mu$, are computed using the approximated value of the $B$ momentum:
\begin{equation}
	\left( p_B \right)_z \simeq \frac{ m_B }{ m_{reco} } \cdot \left( p_{reco} \right)_z,
\label{eq:approx_mu}
\end{equation}
where $m_{reco}$ and $p_{reco}$ are the visible mass and momentum, while $m_B$ is the nominal $B$ mass.
This approximation is valid because of the large boost along the beam momentum ($z$ direction) of the $B$ particles in the LHCb detector.
The yields of signal, normalization and the various background components are obtained through a three-dimensional template fit on $q^2$, $m_{miss}^2$ and $E^*_\mu$, with templates extracted from simulated samples and validated on data-driven control samples (see Figure~\ref{fig:fit_mu}). The result of this measurement is $\mathcal{R}(D^*) = 0.336 \pm 0.027 (\text{stat}) \pm 0.030(\text{syst})$, which is $2.1 \sigma$ larger than the SM expectation.

This document presents the analysis strategy and the perspectives of the measurement of $\mathcal{R} ( D^* )$, using three-prong $\tau$ decays, which has been recently performed at LHCb with data collected during 2011 and 2012 at a centre-of-mass energy of 7 and 8 TeV, corresponding to an integrated luminosity of 3 $\textrm{fb}^{-1}$.

The final part of this proceeding reports the results of the measurement, which were made public after the end of the Conference.

\section{$\mathcal{R}(D^*)$ with three-prong $\tau$ decays}

The signal chosen for the analysis is $B^0 \to D^{*-} \tau^+ \nu_\tau$, where the $D^{*-}$ is reconstructed through the $D^{*-} \to \Dzb( \to K^+ \pi^-) \pi^-$ decay chain, while the $\tau$ lepton is reconstructed through the $\tau^+ \to \pi^+ \pi^- \pi^+ (\pi^0) \neutb$ decay.\footnote{Charge conjugated decay modes are implied throughout the document.}
Since the neutrinos and the $\pi^0$ are not detected, the visible final state consists of five pions plus a kaon.
It is experimentally convenient to measure:
\begin{equation}
	\mathcal{R}_{had} ( D^* ) = \frac{ \mathcal{B}\left( B^{0}\to D^{*-} \tau^{+} \nu_{\tau} \right) }{ \mathcal{B}\left( B^{0}\to D^{*-} \pi^+ \pi^- \pi^+ \right) },
\label{eq:RDstar_had}
\end{equation}
because most of the systematic uncertainties will cancel out in the efficiency ratio, since signal and normalization have the same final state.
Once $\mathcal{R}_{had} ( D^* )$ is measured, it will be multiplied by externally measured branching fractions to obtain $\mathcal{R} ( D^* )$:
\begin{equation}
	\mathcal{R} (D^*) = \mathcal{R}_{had} ( D^* ) \times \frac{\mathcal{B}\left( B^{0}\to D^{*-} \pi^+ \pi^- \pi^+ \right) }{ \mathcal{B}\left( B^{0}\to D^{*-} \mu^{+} \nu_{\mu} \right)}.
\label{eq:RDstar_final}
\end{equation}


The most dominant background consists of inclusive decays of b-hadrons to $D^* 3\pi X$, where the three pions come promptly from the b-hadron decay vertex.
Since the $\tau$ decay vertex is reconstructed with good resolution, it is possible to suppress this kind of background requiring the $\tau$ vertex to be downstream, along the beam direction, with respect to the $B$ vertex with a $4\sigma$ significance.
This selection, applied along with other topological cuts, suppresses the dominant background by three orders of magnitude.

The background surviving the first selection is mainly due to double-charmed $B$ decays, since their topology is very similar to the signal one.
This kind of background is dominated by $B^0 \to D^{*-} D^{+}_s (\to \pi^+ \pi^- \pi^+ X)$ decay, whose branching ratio is 4 times larger than the signal.
In order to discriminate this background from signal, a set of variables is used; they can be grouped in three categories: variables computed with two partial reconstruction techniques, one in signal hypothesis and the other in background hypothesis; isolation variables (i.e. how much the signal tracks are isolated from the other tracks and neutral energy deposits in the event); variables related to the $3\pi$ system dynamics.
These variables are used as input to train a Boosted Decision Tree (BDT).

The partial reconstruction in signal hypothesis allows to compute the squared $B - D^*$ transferred momentum $q^2$ and the $\tau$ decay time with a sufficiently good resolution to maintain separation between signal and background.


Three-dimensional shapes of $q^2$, $\tau$ decay time and BDT output are extracted from simulated and data-driven control samples which represent the various contributions in data.
In order to extract the signal yield, the three-dimensional shapes are used to perform an extended maximum-likelihood template fit on data in high-BDT region.
The various templates used in the fit can be grouped in 5 categories: signal (both $\tau^+ \to \pi^+ \pi^- \pi^+ \neutb$ and $\tau^+ \to \pi^+ \pi^- \pi^+ \pi^0 \neutb$), excited $D^*$ contributions (i.e. $B^0 \to D^{**} \tau^+ \nu_\tau$), double-charmed components, $B^0 \to D^{*-} \pi^+ \pi^- \pi^+ X$ background and combinatorial background.

Since the relative fractions of the various $D^+_s \to \pi^+ \pi^- \pi^+ X$ decays are currently not well known, they are measured in the low-BDT region, which is enriched in such decays and where the signal is absent.
Four different templates in $\min[\pi^+ \pi^-]$, $\max[\pi^+ \pi^-]$, $m(\pi^+ \pi^+)$ and $m(3\pi)$ are built, corresponding to: events where at least one pion comes from an $\eta'$ resonance, events where at least one pion comes from an $\eta$ resonance but none of them originates from an $\eta'$, events where the pions come from a resonance which is not $\eta'$ nor $\eta$ and events where the pions do not originate from a resonance.
A template fit is performed, and the resulting relative fractions from the low-BDT region are then used to constrain the $D^+_s$ decay model in the high-BDT region.

\begin{figure}[htb]
\begin{center}
	\includegraphics[width=0.4\linewidth]{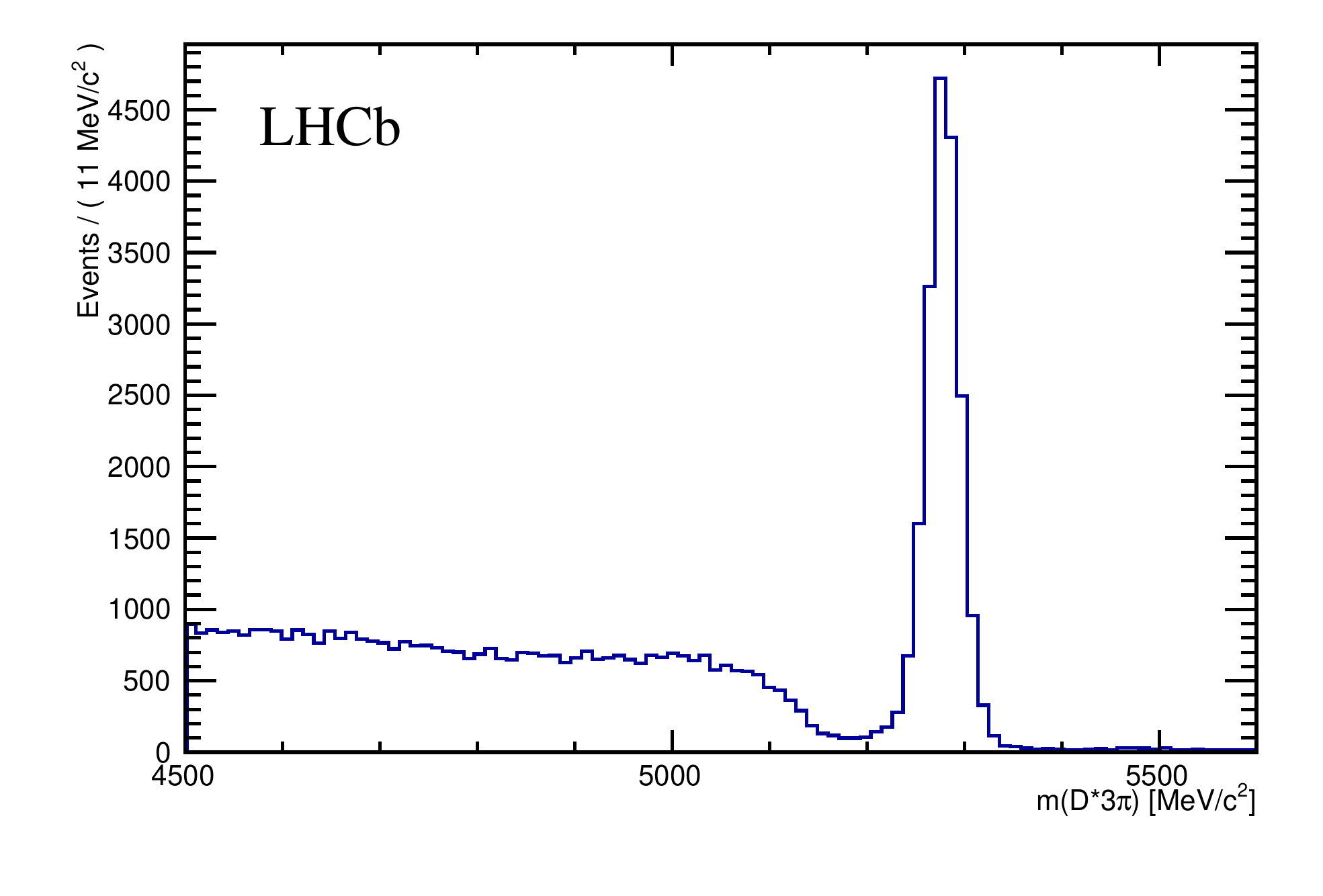}
	\includegraphics[width=0.4\linewidth]{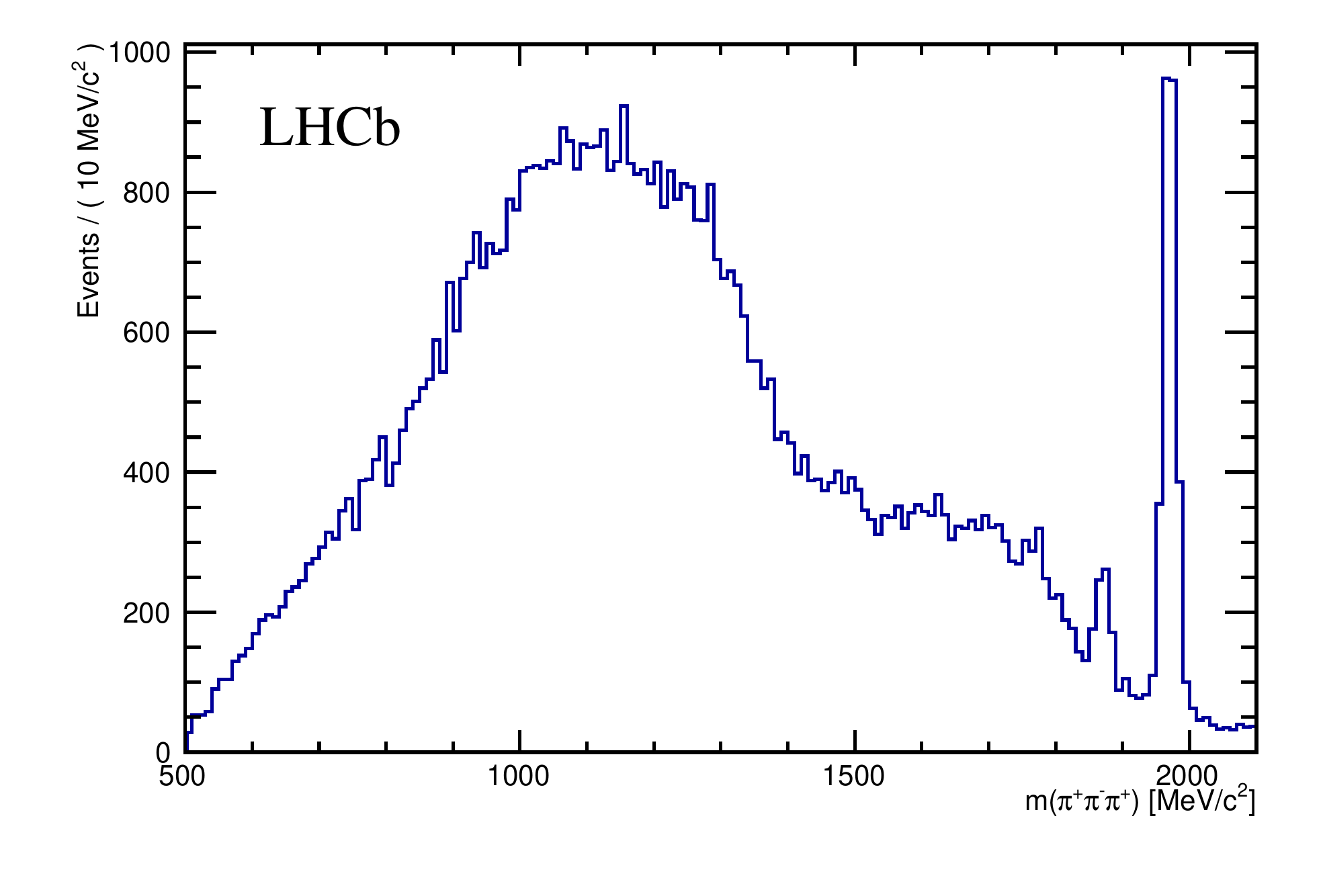}
	\includegraphics[width=0.4\linewidth]{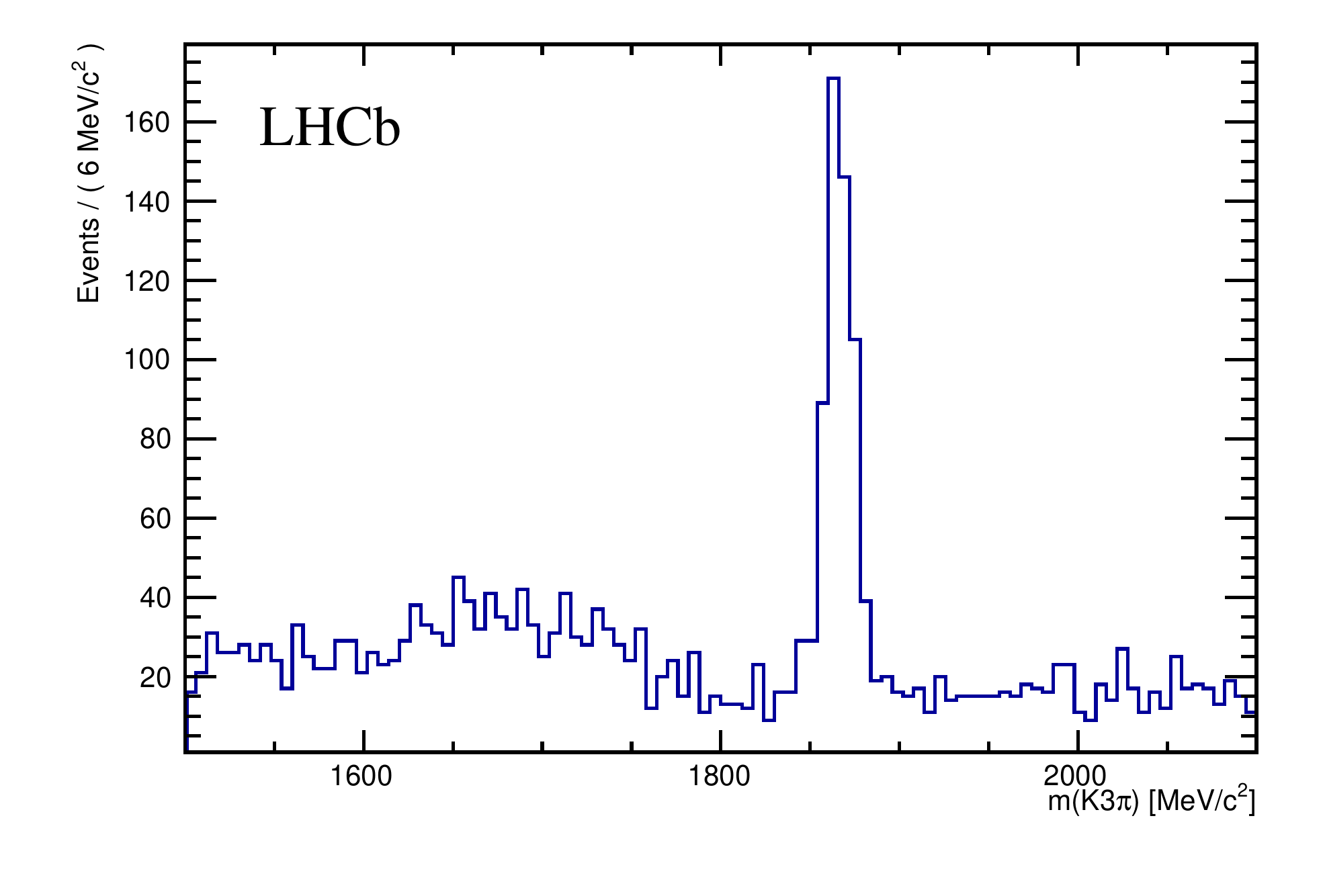}
	\includegraphics[width=0.4\linewidth]{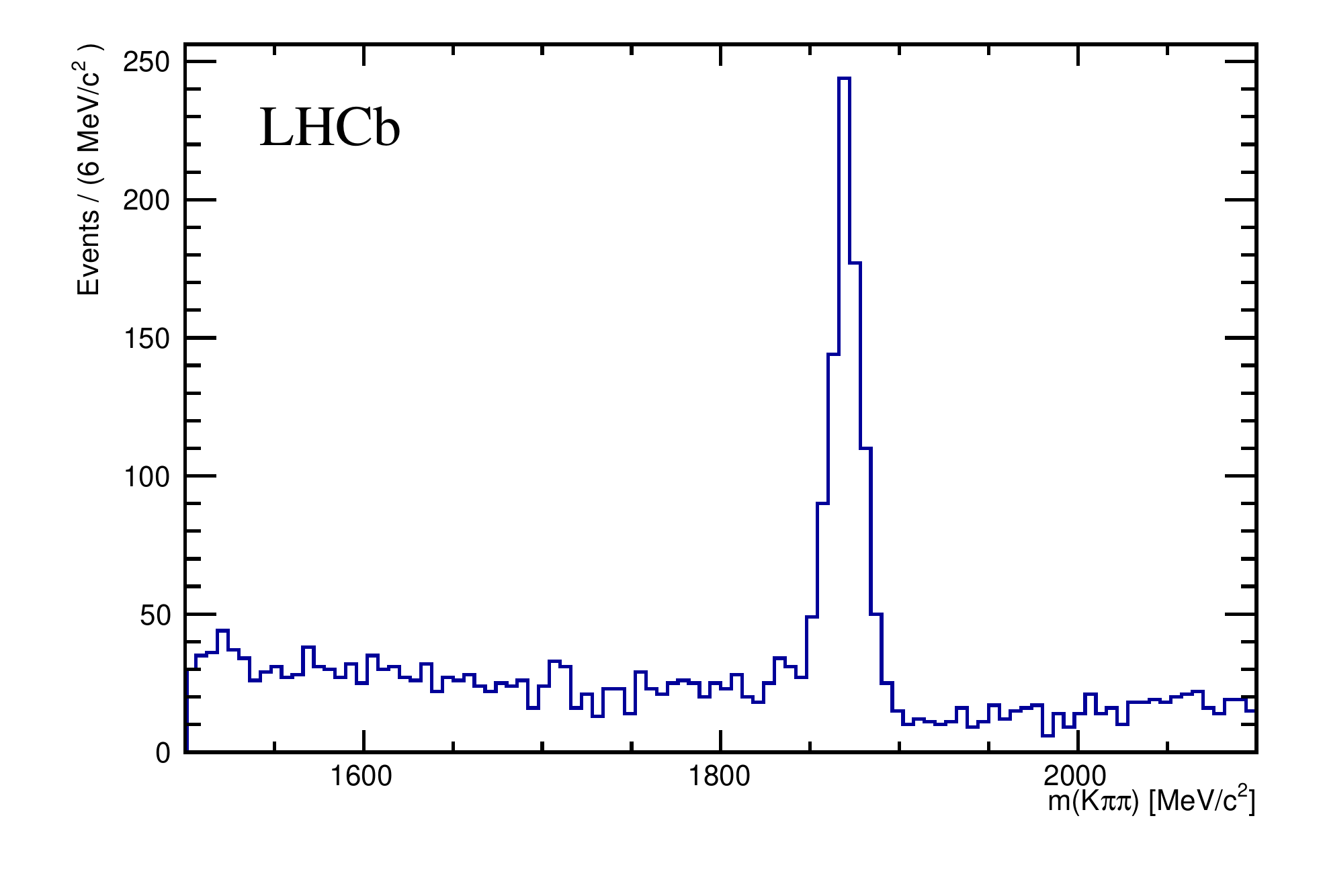}
\end{center}
\caption{
(top left) $D^* 3\pi$ invariant mass distribution for normalization events.
(top right) $\pi^+ \pi^- \pi^+$ invariant mass distribution; the peak in the $D^+_s$ region is used to extract the $B \to D^* D^+_s X$ control sample.
(bottom left) $K^- \pi^+ \pi^- \pi^+$ invariant mass distribution; the peak in the $D^0$ region is used to obtain the $B \to D^* D^0 X$ control sample.
(bottom right) $\pi^+ \pi^- \pi^+$ invariant mass distribution with kaon hypothesis on the $\pi^-$; the peak in the $D^+$ region is needed to build the $B \to D^* D^+ X$ control sample.
}
\label{fig:ctrl_samples}
\end{figure}

To select normalization events, the $\tau$ vertex requirement is reversed, i.e. the $\tau$ vertex is required to be upstream with respect to the $D^0$ vertex with a $4\sigma$ significance.
The normalization yield is obtained by fitting the $D^* 3\pi$ invariant mass distribution (see Figure~\ref{fig:ctrl_samples}) in the $B$ region.
 



In order to validate the simulated samples, three control samples extracted from data are used (see Figure~\ref{fig:ctrl_samples}):
\begin{itemize}
	\item $B \to D^* D^+_s X$ sample, obtained by selecting events in the exclusive $D^+_s \to \pi^+ \pi^- \pi^+$ peak in the $3\pi$ invariant mass distribution.
	\item $B \to D^* D^0 X$ sample, selected by requiring a charged kaon around the $3\pi$ vertex and the $K 3\pi$ invariant mass to be compatible with the $D^0$ mass.
	\item $B \to D^* D^+ X$ sample, obtained by requiring kaon particle identification criteria for the $\pi^-$ in the $\pi^+ \pi^- \pi^+$ system, and the $K^- \pi^+ \pi^+$ to be compatible with the $D^+$ mass.
\end{itemize}

\section{Result and perspectives}

The result of the measurement~\cite{lhcb2017} is:

\begin{equation}
	\mathcal{R}(D^*) = 0.285 \pm 0.019 (\text{stat}) \pm 0.025(\text{syst}) \pm 0.013 (\text{ext}),
\label{eq:result}
\end{equation}

where the third uncertainty originates from the limited knowledge of the branching fraction of the normalization decay mode.
This measurement has the best statistical precision among all the measurements of $\mathcal{R}(D^*)$ performed so far.
The largest systematic uncertainties are due to the limited statistics of the simulated samples and to the precision on the knowledge of the external branching ratios $\mathcal{B}\left( B^{0}\to D^{*-} \mu^{+} \nu_{\mu} \right)$ and $\mathcal{B}\left( B^{0}\to D^{*-} \pi^+ \pi^- \pi^+ \right)$.
Another important source of systematic uncertainty is due to the knowledge of the various $D^+_s$, $D^+$ and $D^0$ background decay models.

\begin{figure}[htb]
	\centering
	\includegraphics[height=3.5in]{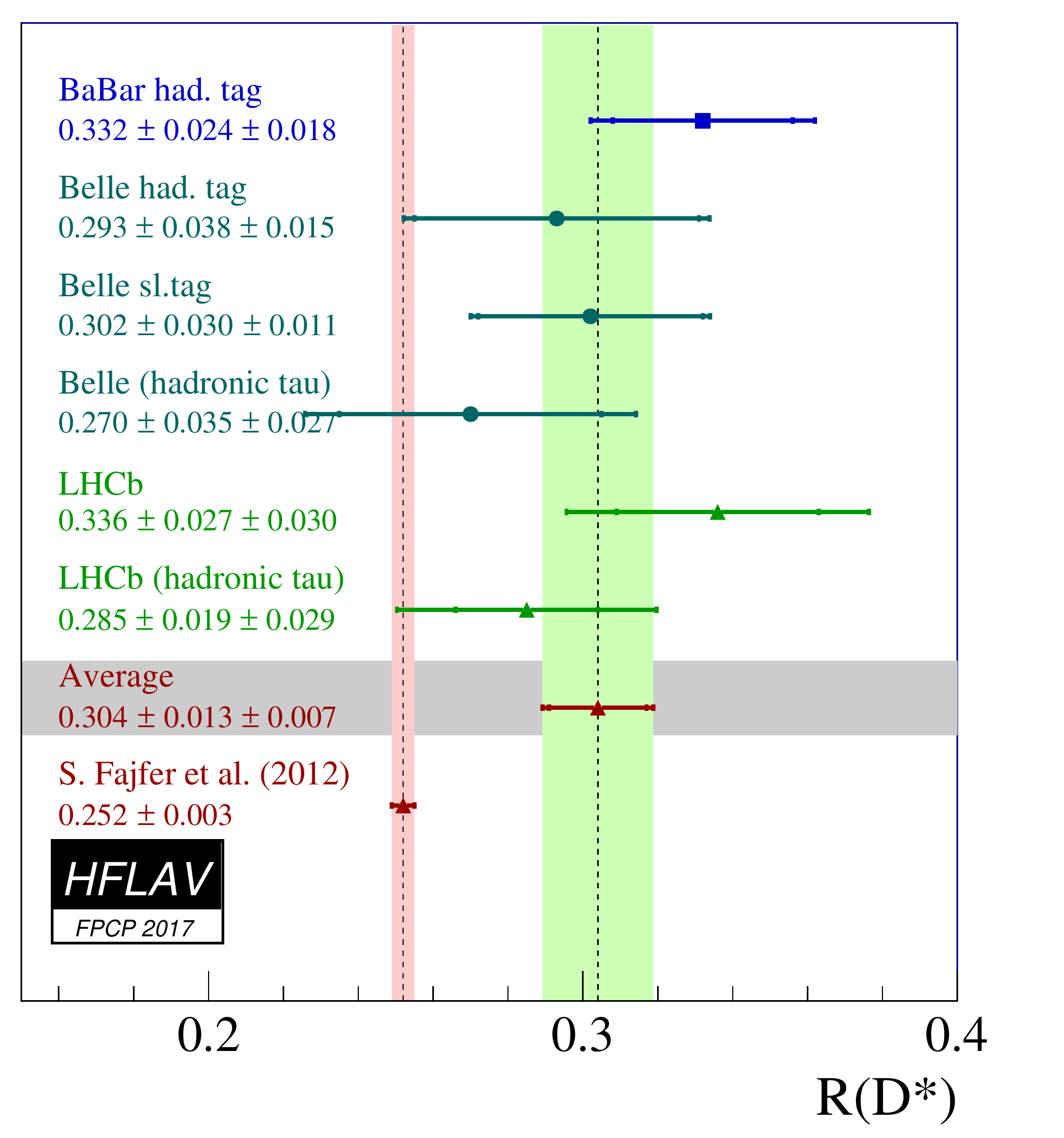}
	\caption{ Current status of $\mathcal{R}(D^*)$ including the measurement with three-prong $\tau$ decays. \cite{HFLAV} }
\label{fig:final_RDstar}
\end{figure}

The value of $\mathcal{R}(D^*)$ obtained from this measurement is higher than the SM calculation and consistent with it within one standard deviation.
An average of this measurement with the LHCb result using $\tau \to \mu \nu_\tau \neumb$, accounting for small correlations due to form factors, $\tau$ polarization and $D^{**}\tau \nu_\tau$ feeddown, gives ${\mathcal{R}}(D^*) = 0.306 \pm 0.016 (\text{stat}) \pm 0.022 (\text{syst})$, consistent with the world average and 2.1 standard deviations above the SM prediction (see Figure~\ref{fig:final_RDstar}). 

%


\end{document}